\documentstyle[12pt]{article}
\def \be{\begin{equation}}
\def \ee{\end{equation}}
\def \bea{\begin{eqnarray}}
\def \eea{\end{eqnarray}}

\newcommand {\bib}[1]   {\bibitem{#1}}

\newcommand {\e}[1]     {\mbox{e}^{#1}}

\newcommand {\yj}       {{x,j}}

\begin{document}

{\large \centerline{\bf Improved Lattice Gauge Field 
Hamiltonian}} 

\vskip 0.3cm

\centerline{Xiang-Qian Luo, and Shuo-Hong Guo}
 
\vskip 0.3cm 

\begin{flushleft}
{\it CCAST (World Laboratory), P.O. Box 8730, Beijing 100080, China,\\
Department of Physics, Zhongshan University, 
Guangzhou 510275, China\footnote{Mailing address} ,\\ 
Center for Computational Physics, School of Physics Science and Engineering,
Zhongshan University, Guangzhou 510275, China, \\
Email: stslxq@zsulink.zsu.edu.cn } 
\end{flushleft}

\vskip 0.3cm

\centerline{Helmut Kr\"oger}
 
\vskip 0.3cm 

\begin{flushleft}
{\it D\'epartement de Physique, Universit\'e  Laval,
Qu\'ebec, Qu\'ebec G1K 7P4, Canada, \\
Email: hkroger@phy.ulaval.ca } 
\end{flushleft}

\vskip 0.3cm 

\centerline{Dieter Sch\"utte}
 
\vskip 0.3cm 

\begin{flushleft}
{\it Institut f\"ur Theoretische Kernphysik,
Universit\"at Bonn, D-53115 Bonn, Germany, \\
Email: schuette@pythia.itkp.uni-bonn.de}
\end{flushleft} 
 
\vskip 0.5cm

\centerline{\bf Abstract}
Lepage's improvement scheme is a recent major progress 
in lattice $QCD$,
allowing to obtain continuum physics 
on very coarse lattices.
Here we discuss improvement
in the Hamiltonian formulation, 
and we derive an improved Hamiltonian 
from a lattice Lagrangian free of $O(a^2)$ errors.
We do this by the transfer matrix method,
but we also show that the alternative via 
Legendre transformation gives identical results.
We consider classical improvement, 
tadpole improvement and also the structure
of L\"uscher-Weisz improvement.
The resulting color-electric energy
is an infinite series, which 
is expected to be rapidly convergent.
For the purpose of practical calculations,
 we construct a simpler 
improved Hamiltonian, which includes only
nearest-neighbor interactions.

\newpage

\section{\bf Introduction}\label{Intro}
Lepage's improvement scheme\cite{Lepage96} is a recent major progress 
in lattice $QCD$,
opening the possibility to approach continuum physics 
on very coarse lattices.
In this paper we want to address the question
how this improvement scheme can be formulated
for a lattice gauge theory in the Hamiltonian approach. 
Although standard lattice gauge theory has been very successful over the last 
two decades, there are areas where progress has been quite slow. Examples are the 
dynamical computation of the $S$-matrix and cross sections, $QCD$ at finite baryon density, or the computation of $QCD$ structure functions in the region of small $x_{B}$ and $Q^{2}$. This situation calls for the development of new methods, and in our opinion 
the lattice Hamiltonian approach is a viable alternative \cite{Kroger92} which should be explored.
The Hamiltonian approach corresponds to consider a continuous time, i.e. $a_t=0$.
Similar ideas have been pursued recently by workers in standard lattice gauge theory 
by considering anisotropic lattices with lattice spacings $a_t << a_s$. 
This has the purpose to improve the computation of the mass spectrum \cite{Lepage96,Morningstar97}.

\bigskip

As a first step, we here restrict ourselves to
the problem of improvement in pure gauge theory.
Let us recall that the improvement program
by Lepage consists of several steps,
by starting from the Wilson action:
First defining a classically improved
action, second performing tadpole improvement
and third introducing 
additional quantum corrections
(L\"uscher-Weisz improvement).
We now discuss how to
carry over these ideas to the
Hamiltonian formulation.
Different strategies are possible,
let us explain those for the case 
of classical improvement.
\newline
\noindent Strategy 1:
Construct the classical Hamiltonian corresponding
to the classical Wilson action. Improve
this classical Hamiltonian and quantize
this Hamiltonian according to the 
rules of canonical quantization.
This yields a {\it classically improved
quantum Hamiltonian}.
\newline
\noindent Strategy 2:
Starting from the classically improved Wilson
action, construct via the transfer matrix
a classically improved quantum Hamiltonian.
\newline
\noindent Strategy 3:
Starting from the classical Wilson action, 
construct first the corresponding 
quantum Hamiltonian via the transfer matrix.
This yields the Kogut-Susskind Hamiltonian
$H(E,U)$ where $U$ and $E$ are the link variables
and their canonical conjugate momenta. 
The usual expansion in powers of the
field variables $A$ and their conjugate 
$\partial/\partial A$ yields then 
the standard expression
$\int d^3x ((\partial/\partial A)^2 + B^2(A))$ 
up to correction of $(a^2)$. 
This Kogut-Susskind Hamiltonian may be improved 
by adding corrections such that 
an better agreement with this {\em formal} continuum
limit operator is obtained.

\bigskip

Different strategies exist also for the
further improvements (tadpole and L\"uscher-Weisz)
with respect to the quantum Hamiltonian.
In principle, the construction
of these quantum corrections should
start from new perturbative calculations
in the Hamiltonian framework.
The coefficients of a fully improved action
as given by Lepage can only be used as 
starting point for the transfer matrix to obtain
a fully improved Hamiltonian 
if the action is expressed
on a lattice with the time spacing being 
much smaller than the spatial spacing 
($a_t<<a_s$).

\bigskip

In this paper we restrict ourselves
mostly to a discussion of classical
improvement. 
As a first result we show explicitly that
the first strategy - the canonical 
quantization of a classical lattice gauge
theory - is a viable alternative to the
second strategy -  using the transfer matrix - 
leading to the same quantum Hamiltonian, but in a more direct way.

\bigskip

The classically improved quantum Hamiltonian 
obtained in this way has the mathematical structure
of a kinetic part with an infinite number of terms.
The reason for this structure
is given by the fact that the inversion of a
nearly local matrix leads to a non-local
matrix. This being an undesirable feature from the point of
view of practical calculations, we show
that it is possible to use the non-uniqueness of
the improved action to obtain
an improved quantum Hamiltonian containing
only nearest neighbor interaction terms.

\bigskip

Finally we discuss the structure of the
quantum Hamiltonian related to the tadpole
and L\"uscher-Weisz corrections. The determination
of the corresponding coefficients, however,
will be deferred to a future investigation.
Although we do not discuss the third strategy in detail,
a fully consistent computation of the quantum
corrections should eventually lead to the same
result as the first two strategies. 
Note that we discuss here only the improved
Hamiltonian for the purpose to compute
the spectrum. 
The general  existence of such improved
Hamiltonians is discussed in
standard many-body theory in the
context of model space calculations
(\cite{Bloch58,Luscher84}).
Like in the action formulation
general observables require particular
improved operators.

\bigskip

Calculations of the glueball spectrum 
using the coupled cluster method based on the standard Kogut-Susskind Hamiltonian have been done by Luo et al. \cite{Luo96} and Sch\"utte et al. \cite{Schutte97}.
An incorporation of an improved Hamiltonian should be possible 
and one would expect reliable results already in lower order 
of the coupled cluster truncation compared to the standard Kogut-Susskind Hamiltonian. Calculations
in this direction are under way\cite{Neff}.

\section{From Wilson action to Kogut-Susskind Hamiltonian}\label{KogutSusskind} 
\subsection{Canonical method via Legendre transformation}\label{KSLegendre}
Before deriving the improved Hamiltonian, we 
describe in a pedagogical manner how to obtain the
standard Kogut-Susskind lattice Hamiltonian \cite{Kogut75} 
from the classical lattice Lagrangian
using the Legendre transformation \cite{Kogut83,Guo87,Luo90},
and canonical quantization.
Wilson`s Euclidean lattice action is given by ($a \equiv a_s$)
\be
S_{E} = 
\frac{a}{a_{t}} \frac{2N_{c}}{g^{2}} \sum_{t-\Box} 
(1 -  P_{\Box} ) 
+
\frac{a_t}{a} \frac{2N_{c}}{g^{2}} \sum_{s-\Box} 
(1 - P_{\Box} ).
\label{Wilsonaction}
\ee
Here the notation of Ref.\cite{Creutz83} has been used,
i.e., $t-\Box$ stands for time-like and $s-\Box$ for 
space-like plaquettes, respectively, and 
\be
P_{\Box} = \frac{1}{N_{c}} Re ~ Tr(U_{\Box}).
\ee
For later use we need to distinguish between the Euclidean and Minkowski action as well as Lagrangian. Its relation is defined, when going from Minkowski to 
Euclidean time 
by the transformation $it \to t$, $\exp[ i S_{M}] \to \exp[ - S_{E}]$,
$L_{M} \to - L_{E}$ and $S_{M} = \int dt L_{M}$ as well as $S_{E} = \int dt L_{E}$. 
Thus the Euclidean lattice Lagrangian is the following
\bea
L_{E} &=& - \frac{2N_{c}a}{g^{2} a_{t}^{2}} \sum_{x,i} (P_{io} -1) 
- \frac{2N_{c}}{g^{2} a} \sum_{x,i<j} (P_{ij} -1) 
\nonumber \\
&=& -{a \over g^2 a_{t}^{2}} \sum_{x,i} 
Tr(U_{i0}(x) + U_{i0}^{\dagger}(x)-2)
\nonumber \\
&-& {1 \over g^2 a} \sum_{x,i<j} 
Tr(U_{ij}(x) + U_{ij}^{\dagger}(x)-2).
\label{Eucllagr}
\eea
Here $U_{ij}(x,t)$ denotes a space-like plaquette where the first link goes in direction $i$ and the second link goes in direction $j$, and $U_{i0}(x,t)$ denotes the corresponding time-like plaquette.  
One should note that $P_{i0}=P_{0i}$ and $P_{ij}=P_{ji}$ for plaquettes.
In the temporal gauge $U_0(x,t) = 1$, the time-like plaquette
becomes a function only of the link variables
$U_i(x,t)$ ($i=1,2,3$) and we have
\be 
P_{i0} = \frac{1}{N_{c}} Re ~ Tr(U_{i0})
= \frac{1}{N_{c}} Re ~ Tr \left( U_i(x,t)U^\dagger_i(x,t+a_t) \right).
\ee
We want to construct a classical Lagrangian 
defining trajectories of generalized coordinates 
$U_i(x,t)$ and generalized velocities, where the
variable $t$ is now continuous.
We assume that the action corresponding
to this Lagrangian is given by the
continuum limit $a_t\to 0$ of the lattice action
which also yields the dependence on
the generalized velocities.

\bigskip

In order to construct this Lagrangian, we introduce 
a Taylor expansion in time
and write for a fixed $(i,x)$
up to errors of $O(a_t^3)$
\begin{eqnarray}
P_{i0} -1 &=& \frac{1}{N_{c}} Re ~ Tr[U(t) U^{\dagger}(t+a_t)-1] 
\nonumber \\
&=& \frac{1}{N_{c}} Re ~ Tr \left( U(t)[U^{\dagger}(t) + a_t {\dot U}^{\dagger}(t)
+{a_t^2 \over 2} {\ddot U}^{\dagger}(t)] -1 \right) 
\nonumber \\
&=& \frac{a_t^2}{2 N_{c}} Re ~ Tr [U(t) {\ddot U}^{\dagger}(t)]
= - \frac{a_t^2}{2 N_{c}} Tr [{\dot U}^{\dagger}(t) {\dot U}(t)]
\nonumber \\
&=& - \frac{a_t^2}{2 N_{c}} Tr [{\dot q}(t) {\dot q}(t)].
\label{q2}
\end{eqnarray}
Here, we denote $U(t) \equiv U_i(x) \equiv U_i(x,t)$, 
and we have introduced  a generalized velocity $\dot{q}$ (corresponding
to the angular velocities of the classical top theory)  
which is an element of the $SU(N_c)$ Lie algebra,
\begin{eqnarray}
{\dot q}_i(x)={\dot q}^{\alpha}(x) \lambda^{\alpha}
=-i {\dot U}_i(x) U^{\dagger}_i(x)
=i U_i(x) {\dot U}^{\dagger}_i(x).
\label{velocity}
\end{eqnarray}
The $SU(N_c)$ generators $\lambda^a$ are  normalized to
$tr(\lambda^a\lambda^b) = \delta^{ab}/2$.
Going to the limit $a_t \to 0$ and performing a transition
from Euclidean to Minkowski space ($t \to it$) yields the
classical lattice Lagrangian
\begin{eqnarray}
L_{M}=
{a \over 2 g^2} \sum_{x,i} 
{\dot q}_i^{\alpha} (x){\dot q}_i^{\alpha} (x)
+
{1 \over g^2 a} \sum_{x,i<j} 
Tr(U_{ij} + U_{ij}^{\dagger}-2).
\end{eqnarray}
Here we denote $U_{ij} \equiv U_{ij}(x) \equiv U_{ij}(x,t)$.   
For a classical canonical formulation
we introduce the conjugate momenta
\begin{eqnarray}
E_j^{\alpha} (x)
&=&{\partial L_{M} \over \partial  {\dot q}_j^{\alpha}(x)}
= {a \over g^2} {\dot q}_j^{\alpha} (x)
={2ai \over g^2} Tr [\lambda^{\alpha} U_j(x) {\dot U^{\dagger}_j(x)}],
\nonumber \\
E_j(x)&=&E_j^{\alpha} (x)\lambda^{\alpha}={a \over g^2} {\dot q}_j(x).
\label{conjmom}
\end{eqnarray}
The standard Legendre transformation 
leads then to the following classical lattice Hamiltonion
\begin{eqnarray}
H&=& \sum_{x,i}
{\partial L_{M} \over \partial {\dot q}_i^{\alpha}(x)}{\dot q}_i^{\alpha}(x)
-L_{M}
\nonumber \\
&=&{g^2 \over 2a} \sum_{x,i} 
 E_i^{\alpha} (x) E_i^{\alpha} (x)
 -
{1 \over g^2 a} \sum_{x,i<j} 
Tr(U_{ij} +U_{ij}^{\dagger}-2).
\label{classicalKS}
\end{eqnarray}
Recalling ${\dot q_i} \to ga{\dot A}$,
we convince ourselves that 
$E_i \approx a^2 {\dot A}_i/g$.
Therefore $E_i(x)$ is the approximated color-electric field 
on the lattice.

\bigskip

To quantize this classical theory,
we proceed according to the rules
of quantization of the classical top theory\cite{Edmonds57}.
This results in the prescription
that the quantum mechanical states 
are functions of the link variables
$U_i(x)$ and that the canonical conjugates
$E_i(x)$ become the operators
of infinitesimal left multiplication.
A generalization of the
standard quantum mechanical formula
$x + a = \e{-i\hat{p}a} x \e{i\hat{p}a}$
yields for the link variables
\be
\e{i \epsilon^{\alpha} \lambda^{\alpha}}  U_j(x)
=
e^{-i \epsilon^{\alpha} E_j^{\alpha} (x)} 
U_j(x) e^{i \epsilon^{\alpha}  E_j^{\alpha} (x)}.
\ee
Variables corresponding to different lattice links
are considered to be independent.
This yields the commutation relations
\begin{eqnarray}
\lbrack U_i(x), E_j^{\alpha} (y)\rbrack
&=&\lambda^{\alpha} U_i(x)\delta_{x,y} \delta_{i,j},
\nonumber \\
\lbrack U^{\dagger}_i(x), E_j^{\alpha} (y)\rbrack
&=&-U^{\dagger}_i(x) \lambda^{\alpha} \delta_{x,y} \delta_{i,j}.
\label{commutator}
\end{eqnarray}
Since the operators $\e{iE^{\alpha}_j(x) \lambda^{\alpha}}$ yield a representation
of the gauge group $SU(N_c)$, we obtain for the 
$E_j(x)$ the commutation relations of the Lie algebra
\be
[E^{\alpha}_j(x),E^{\beta}_j(x)] = 
i f^{\alpha \beta \gamma} E^{\gamma}_j(x).
\ee
The quantization of the classical Hamiltonian, Eq.(\ref{classicalKS}),
by use of the commutation relations Eq.(\ref{commutator})
gives the standard quantum Hamiltonian of Kogut and Susskind.

\subsection{The transfer matrix method}\label{KStransfer} 
The construction of the Kogut Susskind Hamiltonian
from the Wilson action via the transfer matrix 
method has been first established by Creutz\cite{Creutz83}
(see also Ref.\cite{Montvay94}).
Here we recall the basic steps
which may be used also for the construction
of the improved Hamiltonian as discussed below.
\be
S_{E} = \int dt \; L_{E} = \sum_t a_t [L_0(q(t),q(t+a_t)) + L_1(q(t))]
+ O(a_{t}^{2}).
\ee
Hereby, $L_0$ is kinetic part of the
Lagrangian which couples the system at time 
$t$ to that at time $t+a_{t}$.
Invoking the Baker-Cambell-Hausdorf formula
and going to the limit $a_t \rightarrow 0$ \cite{Montvay94a},
the Hamiltonian is eventually given by
\be
H = H_0 + L_{1},
\ee
where the nontrivial part $H_0$ is related to $L_0$
via the functional integral kernel of the 
corresponding time evolution operator
(transfer matrix formalism).

\bigskip

Let us recall that relation for the 
simple example of standard one-body quantum mechanics of free motion
\cite{Montvay94a} where one has (we put the mass $m=1$)
\be
L_0(q',q) = \frac{1}{2 a_{t}^2} (q' - q)^2,
\label{QML0}
\ee
with $ q' = q(t+ a_{t})$, $q = q(t)$.
The discrete time-evolution, which relates the generator $H_0$ to the kernel 
$L_{0}$, is given by 
\be
(\e{-a_{t} H_0}\psi)(q) = N\int dq' \e{-a_{t} L_0(q,q')} \psi(q'),
\ee
where $N$ is some unimportant normalization factor.
Using
\be
\psi(q') = \e{(q'-q)\nabla}\psi(q),
\label{evolQMwfct}
\ee
yields
\be
\e{-a_{t} H_0} =N \int dq' \e{-a_{t} L_0(q,q')} \;\e{(q'-q)\nabla}.
\label{evolQMHam}
\ee
In this case, and also for the case of scalar
field theory \cite{Montvay94a},
this integral is analytically computable 
for finite $a_{t}$. It yields the usual
result $H_0 = -\Delta/2$.

\bigskip

In order to clarify the notations and the 
particularities for non-abelian gauge field
theories, we first recall how to obtain from the transfer 
matrix method the standard Kogut Susskind Hamiltonian.
We start by the decomposition of the action as given by Eq.(\ref{Wilsonaction})
and introduce the temporal gauge. The corresponding Lagrangian is given by
\bea
L_0(U(t+a_t),U(t)) &=& 
{a \over g^2 a_{t}^{2}} \sum_{x,i} 
Tr(2 -U_{i}(x,t+a_{t}) U^{\dagger}_{i}(x,t) - h.c. ),
\nonumber \\
L_1(U(t)) &=& \frac{1}{g^{2} a} \sum_{x,i<j} Tr(2 - U_{ij}(x,t) - h.c.). 
\eea 
Here $L_{0}$ corresponds to the kinetic part and $L_{1}$ to the potential part, respectively. The kinetic part of the Lagrangian 
is given by the plaquettes
involving different times.
Using the notation analogous to Eq.(\ref{QML0}), 
\bea
L_0(U'_{x,j},U_{x,j}) &=& \frac{a}{g^{2} a_{t}^{2}}
\sum_\yj Tr \left[ 2 - (V_\yj + V_\yj^{\dagger}) \right],
\nonumber \\
V_{x,j} &=& U'_{x,j} U_{x,j}^{\dagger},
\label{L0V}
\eea
where $U$ corresponds to the time slice $t$ and $U'$ to the time slice $t+a_{t}$, respectively.
It is well known that the quantum mechanics of $SU(N_{c})$ gauge 
theory and that of the quantum mechanical top are closely related \cite{Kogut75,Edmonds57}.
Thus the relation between the wave function at time slice $t$ 
and time slice $t+a_{t}$, in analogy to Eq.({\ref{evolQMwfct}), 
involves the standard color electric operators 
$E^{\alpha}, \alpha =1,..,N_{c}^2-1$,
\be
\Psi(U') = \e{i \omega^{\alpha} E^{\alpha}} \Psi(U).
\label{evolgaugewfct}
\ee
where the numbers $\omega^{\alpha}$ are the parameters of 
a group element $g_{V} \in SU(N_c)$ such that
\be
g_{V} = \e{i \omega^{\alpha} \lambda^{\alpha} } = U' U^{-1} = V.
\ee
The analogy to Eq.(\ref{evolQMHam}) the Hamiltonian $H_0$
in the case of one link variable $V = U'U^\dagger$ is given by
\be
\e{-a_{t} H_0} = N \int dU'\e{-a_{t} L_0(V)} 
\e{i \omega^{\alpha} E^{\alpha}},
\label{evolgaugeHam}
\ee
where $L_{0}(V)$ is given by Eq.(\ref{L0V}).
We use the invariance of the Haar measure 
yielding $dU' = dV$ 
and note that like in 
Eqs.(\ref{evolQMwfct},\ref{evolgaugewfct}), the
operators $E^a$ have to be treated as commuting with
$U$ and $U'$.
The integral in Eq.(\ref{evolgaugeHam}) can not be 
evaluated analytically for finite time translations $a_{t}$.
However, for the determination of $H_0$
one may consider the limit $a_{t} \rightarrow 0$.
In this case, the variables $V$ approach the
identity  and it is legitimate to
use the approximation for the $L_{0}$ term
\be
Tr(V + V^{\dagger}) = 2 Tr(\cos \lambda^{\alpha} \omega^{\alpha}) = 
2\left[ N_{c} -\frac{1}{4} \omega^{2} + O(\omega^{4}) \right].
\ee
Writing the group integral $\int dU$ as Haar measure 
$\int \prod_a d\omega^a det_{Jac}$ yields 
for Eq.(\ref{evolgaugeHam}) a Gaussian integral in analogy to Eq.(\ref{evolQMHam}), 
with the exponent 
\be
\frac{a}{2 a_{t} g^2} \omega^{\alpha} \omega^{\alpha} + i E^{\alpha} \omega^{\alpha} =
\frac{a}{2 a_{t} g^2} (\omega^{\alpha}  + i E^{\alpha} \frac{ a_{t} g^2}{a})^2 
+ \frac{a_{t} g^2}{2 a} E^{\alpha} E^{\alpha}.
\ee
Taking the sum over the space-like links this
reproduces the standard kinetic term of the Kogut Susskind Hamiltonian
\be
H_0 = \frac{g^2}{2 a} \sum_\yj  E^{\alpha}_{j}(x) E^{\alpha}_{j}(x).
\ee

\section{Classical improvement of Kogut Susskind Hamiltonian}
\subsection{Continuum behavior of classical improved action}
The Wilson action reproduces the classical continuum action only up to errors of $O(a^{2})$. It is possible to add to the Wilson action new terms such that 
these $O(a^{2})$ errors are canceled \cite{Symanzik83,Hasenfratz94,Lepage96}.
In order to construct the corresponding improved Hamiltonian, one needs a generalization to lattices with $a_{t} \neq a_{s} \equiv a$.
We first discuss the classical continuum behavior of the Wilson action.
For space-like plaquettes one has (see Ref.\cite{Lepage96})
\begin{eqnarray}
P_{ij}={1 \over N_c} Re ~ Tr(U_{ij})
\to 1 - {g^2 a^4 \over 2N_c} 
\left( Tr [F_{ij} F_{ij}] +  {a^{2} \over 12}  
Tr [F_{ij}({\cal D}_{i}^2 + {\cal D}_{j}^2) F_{ij}] \right).
\nonumber \\
\end{eqnarray}
For convenience, the continuum limit of a loop is expressed in terms of the
field strength tensor and its covariant derivative the center of the loop $x_{0}$. According to Ref.\cite{Lepage96} for time-like plaquettes one has
to consider the path ordered integral 
\begin{eqnarray}
\oint A \cdot dx 
&\to& 
\int_{-a/2}^{a/2} dx_i \int_{-a_t/2}^{a_t/2} dt
[F_{i0}(x_0) +  {1 \over 2} x_{i'}x_{j'} 
({\cal D}_{i'} {\cal D}_{j'} F_{i0}) \vert_{x=x_0}]
\nonumber \\
&\to& 
a a_t F_{i0}(x_0) + 
{a_t a^3 \over 24} ({\cal D}_{i}^2) F_{i0}(x_0)
+{a a_t^3 \over 24} ({\cal D}_{0}^2) F_{i0}(x_0).
\end{eqnarray}
The last term can be neglected since $a_t << a$.
Therefore
\begin{eqnarray}
P_{i0} 
&=& 
{1 \over N_c} Re ~ Tr(U_{i0})
\to {1 \over N_c} Re Tr [1-{1 \over 2} (g\oint A \cdot dx)^2]
\nonumber \\
&\to& 
{1 \over N_c}  
Re ~ Tr [1-{g^2 a^2 a_t^2 \over 2}
(F_{i0}+{1 \over 24}a^2{\cal D}_{i}^2 F_{i0})^2]
\nonumber \\
&\to& 
1 - {g^2 a^2 a_t^2 \over 2N_c} 
\left( Tr [F_{i0} F_{i0}]
+ {a^2 \over 12} 
Tr [F_{i0} {\cal D}_{i}^2 F_{i0}] \right)
+O(a^2 a_t^4).
\label{ContPi0}
\end{eqnarray}

\bigskip

In order to compensate these $O(a^2)$ errors, Lepage\cite{Lepage96}
has proposed to add new terms to the Wilson action. 
One of these terms is given by a rectangular loop,
\begin{eqnarray}
R_{\mu\nu}= {1 \over N_c} Re ~ Tr
\setlength{\unitlength}{.015in}
\setlength{\unitlength}{.015in}
\begin{picture}(75,13)(0,17)
\put(10,5){\makebox(0,0){$x$}}
  \put(10,30){\vector(0,-1){12.5}}
  \put(10,10){\line(0,1){20}}
  \put(50,30){\vector(-1,0){13}}
  \put(50,30){\vector(-1,0){33}}
  \put(10,30){\line(1,0){40}}
  \put(50,10){\vector(0,1){12.5}}
  \put(50,30){\line(0,-1){20}}
  \put(10,10){\vector(1,0){12.5}}
  \put(10,10){\vector(1,0){32.5}}
  \put(50,10){\line(-1,0){40}}
  \put(61,10){\vector(1,0){10}}\put(75,10){\makebox(0,0){$\mu$}}
  \put(60,11){\vector(0,1){10}}\put(60,25){\makebox(0,0){$\nu$}}
\end{picture}
\end{eqnarray}
For a space-like loop one has   
in particular
\begin{eqnarray}
R_{ij} 
&=& 
{1 \over N_c} Re ~ Tr[U_i(x) U_i(x+a\hat{i}) U_j(x+2a\hat{i}) 
\nonumber \\
&\times& 
U^{\dagger}_i(x+a\hat{j}+a\hat{i})
U^{\dagger}_i(x+a\hat{j}) U^{\dagger}_j(x) ]
\nonumber \\
&\to& 
1 - {g^2 a^4 \over  2N_c} 
\left( 4 Tr [F_{ij} F_{ij}] 
 + {a^2 \over 3}  
Tr [F_{ij}(4 {\cal D}_{i}^2 + {\cal D}_{j}^2) F_{ij}] \right).
\end{eqnarray}
Considering time-like loops, there are two possibilities.
First, one has a $2a \times a_t$  rectangular loop
\begin{eqnarray}
R_{i0}= {1 \over N_c} Re ~ Tr
\setlength{\unitlength}{.015in}
\begin{picture}(75,13)(0,17)
\put(10,5){\makebox(0,0){$x$}}
  \put(10,30){\vector(0,-1){12.5}}
  \put(10,10){\line(0,1){20}}
  \put(50,30){\vector(-1,0){13}}
  \put(50,30){\vector(-1,0){33}}
  \put(10,30){\line(1,0){40}}
  \put(50,10){\vector(0,1){12.5}}
  \put(50,30){\line(0,-1){20}}
  \put(10,10){\vector(1,0){12.5}}
  \put(10,10){\vector(1,0){32.5}}
  \put(50,10){\line(-1,0){40}}
  \put(61,10){\vector(1,0){10}}\put(75,10){\makebox(0,0){$i$}}
  \put(60,11){\vector(0,1){10}}\put(60,25){\makebox(0,0){$0$}}
\end{picture}
\end{eqnarray}
yielding
\begin{eqnarray}
R_{i0} 
&=& 
{1 \over N_c} Re ~ Tr[U_i(x,t) U_i(x+a\hat{i},t) U_0(x+2a\hat{i},t) 
\nonumber \\
&\times& 
U^{\dagger}_i(x+a\hat{i},t+a_{t}) 
U^{\dagger}_i(x,t+a_{t}) U^{\dagger}_0(x,t) ]
\nonumber \\
&\to& 
1- {g^2 a^2 a_t^2 \over 2 N_c} 
Tr Re (2 F_{i0}+ {a^2 \over 3} {\cal D}_{i}^2 F_{i0})^2
\nonumber \\
&\to& 
1 - {g^2 a^2 a_t^2 \over 2N_c} 
\left( 4 Tr [F_{i0} F_{i0}] 
+ {4 a^2 \over 3} Tr [F_{i0} {\cal D}_{i}^2 F_{i0}] \right).
\label{ContRi0}
\end{eqnarray}

\bigskip

Secondly, one has a $2a_t \times a$  rectangular loop

\bigskip

\begin{eqnarray}
R_{0i}= {1 \over N_c} Re ~ Tr
\setlength{\unitlength}{.015in}
\begin{picture}(75,13)(0,17)

\put(10,0){\vector(1,0){13}}
\put(10,0){\line(1,0){20}}

\put(30,0){\vector(0,1){13}}
\put(30,0){\vector(0,1){33}}    
\put(30,0){\line(0,1){40}}

\put(30,40){\vector(-1,0){13}}
\put(30,40){\line(-1,0){20}}

\put(10,40){\vector(0,-1){13}}
\put(10,40){\vector(0,-1){33}}  
\put(10,0){\line(0,1){40}}
 
\put(61,0){\vector(1,0){10}}\put(75,0){\makebox(0,0){$i$}}
\put(60,1){\vector(0,1){10}}\put(60,15){\makebox(0,0){$0$}}
\end{picture}
\end{eqnarray}

\bigskip

This term corresponds to advancing two steps in time direction. 
The conventional transfer matrix corresponds to an advance of 
a single step in time direction. Thus it is not compatible with the 
definition of the transfer matrix. We may disregard this term
because the improvement terms in the Lagrangian are not uniquely
determined\cite{Lepage96}. 
Taking into account only the first term is sufficient.

\bigskip

Therefore, we make the following ansatz for
the classically improved Euclidean lattice Lagrangian \cite{Lepage96}.
\begin{eqnarray}
L_t
&=&
-{2N_c a \over g^2 a_t^2} \sum_{x,i} 
[C_1' {P_{i0}+P_{0i} \over 2} + C_2' R_{i0}] + \mbox{const.}
\nonumber \\
\nonumber \\
&& C_1' = 4/3  \;\;\; \mbox{and}  \;\;\; C_2' =-1/12,
\nonumber \\
\nonumber \\
L_s &=&
-{2N_c \over g^2 a} \sum_{x,i<j} 
[C_1 {P_{ij}+P_{ji} \over 2} +C_2 (R_{ij}+R_{ji})] + \mbox{const.}
\nonumber \\
\nonumber \\
&& C_1=5/3 \;\;\; \mbox{and} \;\;\; C_2=-1/12,
\nonumber \\
\nonumber \\
L_{E} &=& L_t+L_s.
\label{ImprovEuclLagr}
\end{eqnarray} 

\subsection{Improved Hamiltonian via Legendre transformation}\label{ImproveLegendre}
Now we proceed as in sect.(\ref{KSLegendre}) 
to construct a classical Lagrange function in Minkowski space
in terms of the generalized coordinates $U_i(x)$
and the generalized velocities $\dot q_i(x)$ as defined in Eq.(\ref{velocity}).
Working in the temporal gauge and denoting $\bar{U}(t)=U_i(x+a\hat{i},t)$
yields 
\begin{eqnarray}
R_{i0} 
&=& 
\frac{1}{N_c} Re ~ Tr[\bar{U}(t) \bar{U}^{\dagger}(t+a_t) 
U^{\dagger}(t+a_t) U(t)]
\nonumber \\
&\to& 
\frac{1}{N_c} Re ~ Tr \{ \bar{U}(t)[ \bar{U}^{\dagger}(t) +
a_t \dot{\bar{U}}^{\dagger}(t)
+{a_t^2 \over 2} \ddot{ \bar{U}}^{\dagger}(t) ]
\nonumber \\
&\times& 
[U^{\dagger}(t) +a_t {\dot U}^{\dagger}(t)
+ {a_t^2 \over 2} {\ddot U}^{\dagger}(t) ] U(t) \}
\nonumber \\
&=& 
\frac{1}{N_c} Re ~ Tr[
{a_t^2 \over 2} \bar{U}(t) \ddot{ \bar{U}}^{\dagger}(t)
+{a_t^2 \over 2} \ddot{ U}^{\dagger}(t) U(t)
+ a_t^2 \bar{U}(t) \dot{\bar{U}}^{\dagger}(t) 
\dot{U}^{\dagger}(t) U(t)] 
\nonumber \\
\nonumber \\
&+& \mbox{const.}
\nonumber \\
&\to& 
- \frac{1}{N_c} a_t^2 Tr [ 
\frac{1}{2} {\dot q}_i(x+ a\hat{i}) {\dot q}_i(x+a\hat{i}) 
+ \frac{1}{2} {\dot q}_i(x){\dot q}_i(x)
+ {\dot Q}_i(x){\dot q}_i(x+a\hat{i})]
\nonumber \\
&+& \mbox{const.} ,
\label{Qq}
\end{eqnarray}
where we have introduced the variable
\begin{eqnarray}
{\dot Q}_i(x)=U_i(x)^{\dagger} {\dot q}_i(x) U_i(x). 
\label{Q}
\end{eqnarray}
This gives the following classical improved
lattice Lagrangian in Minkowski space 
\be
L_{M}
={a \over g^2} \sum_{x,i} Tr
[(C'_1+2C'_2) {\dot q}_i(x){\dot q}_i(x)
+ 2 C'_2 {\dot Q}_i(x){\dot q}_i(x+a\hat{i})]
- L_{s}.
\label{Lc}
\ee
This Lagrangian can be written in the form
\bea
L_{M} 
&=& 
\frac{1}{2} \frac{a}{g^{2}} \sum_{\sigma, \rho} {\dot q}_{\sigma}(t) 
M_{\sigma \rho}(U(t))
{\dot q}_{\rho}(t) - L_{s} ,
\nonumber \\
&& 
\mbox{where} ~~~ \sigma =(x,i,\alpha ), ~\rho =(y,j,\beta),
\nonumber \\
\nonumber \\
M_{\sigma, \rho}(U(t)) 
&=& 
(C'_{1} + 2C'_{2}) \delta_{\sigma, \rho} 
\nonumber \\
&+&
4C'_{2} \delta(x, y-a\hat{i}) \delta_{i,j} Tr[ U_{i}^{\dagger}(x) \lambda^{\alpha} U_{i}(x) \lambda^{\beta} ].
\label{defmatrixM}
\eea
The matrix $M$ is not symmetric. However, it can be shown that 
only the symmetric part of $M$ will contribute to the Hamiltonian. 
Thus we introduce
\bea
L^{sym}_{\sigma,\rho} &=& 
\delta(x, y-a\hat{i}) \delta_{i,j} Tr[ U^{\dagger}_{x \to y} \lambda^{\alpha} 
U_{x \to y} \lambda^{\beta} ]
\nonumber \\
&+& 
\delta(x, y+a\hat{i}) \delta_{i,j} Tr[ U^{\dagger}_{y \to x} \lambda^{\beta} 
U_{y \to x} \lambda^{\alpha} ],
\label{Lsym}
\eea
which allows to write
\be
M^{sym}_{\sigma,\rho} = \frac{1}{2} (M + M^{t})_{\sigma,\rho}
= 
(C'_{1} + 2C'_{2}) \delta_{\sigma, \rho} 
+
2C'_{2} L^{sym}_{\sigma,\rho}.
\ee
Inspection shows that $L_{sym}$ and hence $M_{sym}$ are real, symmetric matrices.
Then the Lagrangian reads
\be 
L_{M} =
\frac{1}{2} \frac{a}{g^{2}} \sum_{\sigma, \rho} {\dot q}_{\sigma}(t) ~
M^{sym}_{\sigma \rho}(U(t)) ~
{\dot q}_{\rho}(t) - L_{s}.
\ee
Via Legendre transformation, the classical improved
Hamiltonian is obtained
\begin{eqnarray}
H
&=&
\sum_{x,i} {\partial L_{M} \over \partial {\dot q}_i^{\alpha} (x)}
{\dot q}_i^{\alpha}(x)
- L_{M} = H_{0} + V,
\nonumber \\ 
H_{0} &=& \frac{1}{2} \frac{g^{2}}{a} \sum_{\sigma, \rho} E_{\sigma} (M^{sym})^{-1}_{\sigma, \rho} E_{\rho}, 
\nonumber \\
V &=& 
- {2N_c \over g^2 a} \sum_{x,i<j} 
\left[ C_1 {P_{ij}+P_{ji} \over 2} +C_2 (R_{ij}+R_{ji}) \right] .
\label{classimprovHam}
\end{eqnarray}
The color-electric field $E_{\sigma}$ is given by the 
conjugate momentum, 
being related to the generalized velocity $\dot{q}_{\sigma}$ via, 
\be
E_{\sigma} = 
\frac{\partial L } { \partial {\dot q}_{\sigma} }
= \frac{a}{g^{2}} \sum_{\rho} M^{sym}_{\sigma, \rho} \dot{q}_{\rho}.
\ee
The color-electric field,
obeys commutation relations with the link variables
given by Eq.(\ref{commutator}).

\subsubsection{Hopping expansion and algebraic properties of $M_{sym}$}
Taking a closer look to the kinetic part of the improved Hamiltonian reveals that via $M_{sym}^{-1}$ an infinite number of terms enters into the Hamiltonian.
In analogy to the hopping parameter expansion \cite{Montvay94}, which expresses the propagator in terms of powers of a hopping matrix, we introduce $K^{sym}$
\bea
M_{sym} &=& m_{0} \left[ 1 + K_{sym} \right] 
= m_{0}\left[1 + k_{0} L_{sym} \right]
\nonumber \\
m_{0} &=& C_{1}' + 2C_{2}' = \frac{7}{6},
\nonumber \\
k_{0} &=& \frac{2 C_{2}'} { C_{1}' + 2 C_{2}'} = -\frac{1}{7},
\eea
to obtain
\be
M_{sym}^{-1} = \frac{1}{m_{0}} \left[ 1 - K_{sym} + K_{sym}^{2} 
- K_{sym}^{3} + \cdots \right].
\ee
While $K_{sym}$ involves only link variables between next neighbor lattice sites, higher powers of $K_{sym}$ involve links extending over several lattice 
sites. Using the notation
\be
U_{x \to x +(N+1)a \hat{i}} = U_{i}(x) U_{i}(x+a\hat{i}) \cdots U_{i}(x+Na\hat{i}),
\ee
we generalize the definition of $L_{sym}$ to 
\bea
L^{(N) sym}_{\sigma,\rho} &=& 
\delta(x, y - Na\hat{i}) \delta_{i,j} Tr[ U^{\dagger}_{x \to y} \lambda^{\alpha}
U_{x \to y} \lambda^{\beta} ]
\nonumber \\
&+& 
\delta(x, y + Na\hat{i}) \delta_{i,j} Tr[ U^{\dagger}_{y \to x} \lambda^{\beta} 
U_{y \to x} \lambda^{\alpha} ],
\nonumber \\
&N& = 0,1,2,\cdots ,
\label{LNsym}
\eea
where $L^{(0)}_{sym} = 1$, and $L^{(1)}_{sym} = L_{sym}$.
A little algebra shows that the matrix $L^{(N)}_{sym}$ obeys the following product rule
\be
L^{(p)}_{sym} L^{(q)}_{sym} = \frac{1}{2} L^{(p+q)}_{sym} + \frac{1}{2} L^{(|p-q|)}_{sym}.
\ee
Thus we obtain for the lowest powers of $K_{sym}$
\bea
K_{sym} &=& k_{0} L^{(1)}_{sym}, 
\nonumber \\ 
K_{sym}^{2} &=& k_{0}^{2} \left[ \frac{1}{2} L^{(0)}_{sym} + \frac{1}{2} L^{(2)}_{sym} \right],
\nonumber \\ 
K_{sym}^{3} &=& k_{0}^{3} \left[ (\frac{1}{2} + \frac{1}{4}) L^{(1)}_{sym} + \frac{1}{4} L^{(3)}_{sym} \right],
\nonumber \\ 
K_{sym}^{4} &=& k_{0}^{4} \left[ (\frac{1}{4} + \frac{1}{8}) L^{(0)}_{sym} + \frac{1}{2} L^{(2)}_{sym} + \frac{1}{8} L^{(4)}_{sym} \right],
\nonumber \\ 
K_{sym}^{5} &=& k_{0}^{5} \left[ (\frac{1}{2} + \frac{1}{8}) L^{(1)}_{sym} + (\frac{1}{4} + \frac{1}{16}) L^{(3)}_{sym} + \frac{1}{16} L^{(5)}_{sym} \right],
\nonumber \\ 
K_{sym}^{6} &=& k_{0}^{6} \left[ (\frac{1}{4} + \frac{1}{16}) L^{(0)}_{sym} + (\frac{1}{4} + \frac{1}{8} + \frac{1}{16} + \frac{1}{32} ) L^{(2)}_{sym} + (\frac{1}{8} + \frac{1}{16}) L^{(4)}_{sym} + \right.
\nonumber \\
&+& \left. \frac{1}{32} L^{(6)}_{sym} \right],
\nonumber \\
& \vdots &
\eea
It has the general structure
\be
K_{sym}^{n} = k_{0}^{n} \sum_{p=0}^{n} \kappa^{(n)}_{p} L^{(p)}_{sym}.
\label{Ksymmetric}
\ee
The coefficients of lowest order are
\bea
\kappa^{(0)}_{0} &=& 1, 
\nonumber \\
\kappa^{(1)}_{0} &=& 0, ~ \kappa^{(1)}_{1} =1,
\nonumber \\
\kappa^{(2)}_{0} &=& \frac{1}{2}, ~ \kappa^{(2)}_{1} = 0, ~ \kappa^{(2)}_{2} = \frac{1}{2},
\nonumber \\
& \vdots &
\eea
The coefficients $\kappa^{(n)}_{p}$ vanish except when $n$ and $p$ are 
both even or both odd.  
Using Eq.(\ref{Ksymmetric}), we express $M^{-1}_{sym}$ by
\bea
M^{-1}_{sym} &=& \frac{1}{m_{0}} \sum_{p=0}^{\infty} \mu_{p} L^{(p)}_{sym},
\nonumber \\
\mu_{p} &=& \sum_{n=p}^{\infty} (-k_{0})^{n} \kappa^{(n)}_{p}.
\eea
As result, starting from an improved Lagrangian with a 
finite number of terms, one obtains for the improved Hamiltonian 
an expression given by an infinite number of terms.

\bigskip

In the following we will explore more of the algebraic structure of $M_{sym}$
and obtain analytic expressions for the hopping expansion coefficients
$\kappa^{(n)}_{p}$. This will be useful in what follows.
We introduce
\be
J_{\sigma,\rho} =
2 \delta(x, y - a\hat{i}) \delta_{i,j} Tr[ U^{\dagger}_{x \to y} \lambda^{\alpha}
U_{x \to y} \lambda^{\beta} ].
\label{DefJ}
\ee
A little algebra shows that 
\be
J^{n}_{\sigma,\rho} =
2 \delta(x, y - na\hat{i}) \delta_{i,j} Tr[ U^{\dagger}_{x \to y} \lambda^{\alpha}
U_{x \to y} \lambda^{\beta} ],
\label{DefJn}
\ee
and
\be
J J^{t} = J^{t} J =1,
\label{Jorthogonal}
\ee
i.e., $J$  is a real, orthogonal matrix.
Comparison with Eqs.(\ref{Lsym}, \ref{LNsym}) shows
\bea
L_{sym} &=& \frac{1}{2}( J + J^{t}),
\nonumber \\
L^{(p)}_{sym} &=& \frac{1}{2}( J^{p} + (J^{t})^{p}),
\nonumber \\
K_{sym} &=& \frac{k_{0}}{2}( J + J^{t}),
\nonumber \\
M_{sym} &=& m_{0} \left[ 1 + \frac{k_{0}}{2}( J + J^{t}) \right].
\label{RelationMsymJ}
\eea
Using Eq.(\ref{Jorthogonal}), $M_{sym}$ can be factorized,
\be 
M_{sym} = \frac{m_{0}}{1 + C^{2}} (1 + C J) (1 + C J^{t}),
\label{FactorMsym}
\ee
if $C$ is chosen as solution of 
\be 
k_{0} = \frac{2C}{1+C^{2}}.
\ee
Solutions are $C = -7 \pm 4 \sqrt{3}$. Note that $J$ being a real, orthogonal matrix, which has eigenvalues of modulus one, and $|C| \neq 1$, thus the matrix $M_{sym}$ can be inverted and $M_{sym}^{-1}$ is well defined. 
Moreover, we note that $M_{sym}$ is a positive definite matrix. This can be seen directly from  Eq.(\ref{FactorMsym}), which factorizes $M_{sym}$ into a matrix times its Hermitian conjugate.
Also, a lower bound can be estimated using Eq.(\ref{RelationMsymJ}).
$J$ being orthogonal implies $||J|| =1$. Thus $R_{J}$ defined by 
$R_{J}= \frac{1}{2}(J + J^{t})$, being a real, symmetric matrix like $M_{sym}$,
obeys $||R_{J}|| \leq 1$.
Then an arbitrary state vector $\phi$ of unit norm
yields $|<\phi | R_{J} | \phi> | \leq 1$.
Then Eq.(\ref{RelationMsymJ}) implies
\bea
<\phi | M_{sym} | \phi > &=& m_{0} + m_{0} k_{0} < \phi | R_{J} | \phi >
\nonumber \\
&=& \frac{7}{6} - \frac{1}{6} <\phi | R_{J} | \phi > ~ \geq 1,
\eea
showing also that $M_{sym}$ is positive.
To summarize the properties of $M_{sym}$, this is a real, symmetric, positive definite and non-singular matrix. This property is needed for the construction 
of the Hamiltonian via the transfer-matrix, in particular for doing the Gaussian integral.

\bigskip

Factorization of $M_{sym}$, via Eq.(\ref{FactorMsym}), allows to express the kinetic energy term $H_{0}$ of the Hamiltonian, Eq.(\ref{classimprovHam}), as follows
\bea 
H_{0} &=& \frac{g^{2}}{2 a} \frac{1 + C^{2}}{m_{0}} \sum_{\nu} 
\left[ \sum_{\rho} (1 + C J)^{-1}_{\nu \rho} E_{\rho} \right]^{2} 
\nonumber \\
&=& \frac{g^{2}}{a} 
{1+C^2 \over m_0} 
Tr \sum_{x,i} [E_i(x)
-CU_i(x)E_i(x+a\hat{i})  U^{\dagger}_i(x) 
\nonumber \\
&+&
C^2 U_i(x)U_i(x+a\hat{i}) E_i(x+2a\hat{i}) U^{\dagger}_i(x+a\hat{i}) U^{\dagger}_i(x) - \cdots ]^{2}.
\eea
Note that this is an expansion in terms of $C$ and $J$.

\bigskip

Analytic expressions for the coefficients of the hopping expansion can be obtained in the following way:
\bea 
K_{sym}^{n} &=& (\frac{k_{0}}{2})^{n} ( J + J^{t})^{n}
\nonumber \\
&=& (\frac{k_{0}}{2})^{n} \sum_{p=0}^{n} 
\left( 
\begin{array}{c}
n \\ p
\end{array} 
\right)
J^{p} (J^{t})^{n-p}.
\eea
Because this expression is a symmetric matrix and making use of Eqs.(\ref{Jorthogonal},\ref{RelationMsymJ}), one may write
\be
K_{sym}^{n} = (\frac{k_{0}}{2})^{n} \sum_{p=0}^{n} \alpha^{(n)}_{p} 
( J^{p} + (J^{t})^{p} ) =  (\frac{k_{0}}{2})^{n} \sum_{p=0}^{n} \alpha^{(n)}_{p} 
2 L^{(p)}_{sym}.
\ee
Comparison of coefficients yields
\bea
p &=& 0: ~ \alpha^{(n)}_{0} = \frac{1}{2} 
\left( 
\begin{array}{c}
n \\ n/2
\end{array} 
\right)
~ \begin{array}{l}
\mbox{if n is even}, \\  \mbox{zero else},
\end{array} 
\nonumber \\
p &\geq& 1: ~ \alpha^{(n)}_{p} =  
\left( 
\begin{array}{c}
n \\ (n+p)/2
\end{array} 
\right)
~ \begin{array}{l}
\mbox{if n, p are both even or both odd}, \\  \mbox{zero else}.
\end{array} 
\nonumber \\
\eea
Comparison with Eq.(\ref{Ksymmetric}) eventually yields for the hopping expansion coefficients $\kappa^{(n)}_{p}$ the following expression,
\be 
\kappa^{(n)}_{p} = 2^{-n+1} \alpha^{(n)}_{p}.
\ee

\subsection{Improved Hamiltonian via transfer matrix}\label{ImprovedTransfer}
We start from the classically improved Euclidean Lagrangian, given by Eq.(\ref{ImprovEuclLagr}). It is built from space-like plaquettes $P_{ij}$, time-like plaquettes $P_{i0}$ and corresponding rectangular loops $R_{ij}$ and $R_{i0}$. We now want to show that the transfer matrix method yields the same Hamiltonian as has been obtained in the previous section via Legendre transformation. Let us consider the time-like part of the Lagrangian, which yields the kinetic part of the Hamiltonian. The space-like part yields the potential part in a trivial way.  
Using the temporal gauge, one has
\bea
P_{i0} &=& 
\frac{1}{2 N_{c}} Tr \left[ U_{i}(x,t) U^{\dagger}_{i}(x,t+a_{t}) 
+ U_{i}(x,t+a_{t}) U^{\dagger}_{i}(x,t) \right]
\nonumber \\
&=& \frac{1}{2 N_{c}} Tr \left[ V_{i}(x,t) + V^{\dagger}_{i}(x,t) \right],
\eea
using the notation $V_{i}(x,t) = U_{i}(x,t+a_{t}) U^{\dagger}_{i}(x,t)$.
Similarly one obtains for the rectangular loop
\bea
R_{i0} 
&=& \frac{1}{2 N_{c}} Tr \left[ V_{i}(x,t) U_{i}(x,t) V_{i}(x+a\hat{i},t) 
U^{\dagger}_{i}(x,t) \right.
\nonumber \\
&+& 
\left. U_{i}(x,t) V^{\dagger}_{i}(x+a\hat{i},t) U^{\dagger}_{i}(x,t) V^{\dagger}_{i}(x,t) \right].
\eea
The Hamiltonian is defined via the transfer matrix 
like in Eq.(\ref{evolgaugeHam}). Because we consider $a_{t} \to 0$, 
the group integral will be dominated by group elements of $SU(N_{c})$
in the neighborhood of the unit element. 
Thus one can expand the group elements $V_{i}(x,t)$ in a Taylor series of
the Lie group parameters $\omega^{\alpha}_{x,i}(t)$,
\be
V_{i}(x,t) = \exp[ i \omega_{x,i}(t) ]
= 1 + i \omega_{x,i}(t)  
- \frac{1}{2} \omega_{x,i}(t)^{2} + O(\omega^{3}), 
\ee
where we denote 
$\omega_{x,i}(t) = \sum_{\alpha} \omega^{\alpha}_{x,i}(t) \lambda^{\alpha}$.
Thus we arrive at 
\be
Tr[ V_{i}(x,t) + V^{\dagger}_{i}(x,t) ]
= 2 N_{c} -\frac{1}{2} \sum_{\alpha} \omega^{\alpha}_{x,i}(t)^{2} +O(\omega^{3}),
\ee
and hence in the notation of Eq.(\ref{defmatrixM}),
\be
\sum_{x,i} Tr[ V_{i}(x,t) + V^{\dagger}_{i}(x,t) ]
= -\frac{1}{2} \sum_{\sigma \rho} \omega_{\sigma}(t) ~ \delta_{\sigma \rho} ~ \omega_{\rho}(t) + O(\omega^{3}) + \mbox{const}.
\ee
Carrying out the corresponding steps for the rectangular term, one obtains
\bea
&& \sum_{x,i} Tr \left[ V_{i}(x,t) U_{i}(x,t) V_{i}(x+a\hat{i}),t) 
U^{\dagger}_{i}(x,t) \right.
\nonumber \\
&+& 
\left. U_{i}(x,t) V^{\dagger}_{i}(x+a\hat{i}),t) U^{\dagger}_{i}(x,t) V^{\dagger}_{i}(x,t) \right]
\nonumber \\
&=& - \sum_{\sigma \rho} \omega_{\sigma}(t) 
\left[ \delta_{\sigma \rho} + \frac{1}{2} J_{\sigma \rho} \right] 
\omega_{\rho}(t) + O(\omega^{3}) + \mbox{const},
\eea
where $J$ is given by Eq.(\ref{DefJ}).
Inserting this into the time-like Lagrangian $L_{t}$ yields
eventually
\bea
L_{t} &=&
\frac{1}{2} \frac{a}{a_{t}^{2} g^{2}} \sum_{\sigma \rho} 
\omega_{\sigma}(t) \left[ (C'_{1} + 2C'_{2}) \delta_{\sigma \rho} + 4C'_{2} J^{sym}_{\sigma \rho} \right] \omega_{\rho}(t).
\eea
One should note that the matrix $J$ is not symmetric. However, to
the Lagrangian only the symmetric part $J_{sym}=(J+J^{t})/2$ contributes.
Note further that
$J_{sym}= L_{sym}$ and $M_{sym} = (C'_{1}+2C'_{2}) 1 + 4C'_{2}J_{sym}$,
being real, symmetric matrices.
Thus we arrive at
\be
L_{t} =
\frac{1}{2} \frac{a}{a_{t}^{2} g^{2}} \sum_{\sigma \rho} 
\omega_{\sigma}(t) M^{sym}_{\sigma \rho} \omega_{\rho}(t).
\ee
The transfer matrix is then given by
\bea
&& \exp[ - a_{t} H_{0} + O(a_{t}^{2}) ] 
\nonumber \\
&=& \int [\prod_{x,i} d V_{i}(x) ] \exp[ -a_{t} L_{t}(V_{i}(x)) ]
\exp[ i \sum_{x,i,\alpha} \omega^{\alpha}_{x,i} E^{\alpha}_{i}(x) ] 
\nonumber \\
&=& \int [ \prod_{\sigma} d \omega_{\sigma} det_{Jac} ]
\exp \left[ - \frac{1}{2} \frac{a}{a_{t} g^{2}} \sum_{\sigma \rho}
\omega_{\sigma} M^{sym}_{\sigma \rho} \omega_{\rho} 
+ i \sum_{\sigma} \omega_{\sigma} E_{\sigma} \right]
\nonumber \\
&=& N \exp \left[ - \frac{1}{2}  \sum_{\sigma \rho}
E_{\sigma} (\frac{a}{a_{t} g^{2}} M^{sym})^{-1}_{\sigma \rho} E_{\rho} 
\right].
\eea
Thus we obtain
\be
H_{0} = \frac{g^{2}}{2a} \sum_{\sigma \rho}
E_{\sigma}  (M_{sym}^{-1})_{\sigma \rho} E_{\rho}.
\ee
in agreement with the result, Eq.(\ref{classimprovHam}), obtained via Legendre transformation.

\section{\bf Improved Hamiltonian given by finite number of terms}\label{Finitenumberterms}
As was shown in the previous section,
the kinetic energy of 
the classical improved Hamiltonian obtained directly
from Lepage's action is given by an infinitive series of terms. 
Even though the series is rapidly convergent,
such an Hamiltonian 
is too complicated for a practical calculation.
Recalling that the purpose of classical improvement 
is to push the $O(a^2)$ error to order $O(a^4)$, we
show here how to construct a simpler improved Hamiltonian 
corresponding to a finite number of terms
to achieve such a goal. 
In the previous section we have seen that the infinite number of terms in the Hamiltonian
arises due to the inversion of the matrix $M_{sym}$, which itself has only a finite number of terms.
Thus it is plausible that in order to obtain a Hamiltonian given by a finite number of terms, 
one needs to start  from a Lagrangian corresponding to a matrix $M_{sym}$ with an infinite number of terms. Such a construction is possible, because the Lagrangian leading to improvement is not unique.
We start by considering the following type of Wilson loop, which emerges as a generalization 
of the $2a \times a_{t}$ loop $R_{i0}$ to a $(n+1)a \times a_{t}$ loop parallel transporter $R_{ni,0}$ given by

\bigskip

\setlength{\unitlength}{0.012in}
\be
R_{ni,0} = \frac{1}{N_c} Re\;Tr\ \ 
\begin{picture}(130,0)(25,765)
\thinlines
\put( 25,785){\line( 0,-1){ 30}}
\put( 25,755){\line( 1, 0){ 160}}
\put(185,755){\line( 0, 1){ 30}}
\put(185,785){\line(-1, 0){ 30}}
\put(155,785){\line( 0,-1){ 25}}
\put(155,760){\line(-1, 0){ 100}}
\put( 55,760){\line( 0, 1){ 25}}
\put( 55,785){\line(-1, 0){ 30}}
\put(25,739){\makebox(0,0){$x$}}
\put(102,736){\makebox(0,0){$x_0$}}
\put(159,738){\makebox(0,0){$x+na\hat{i}$}}
\nonumber \\
\end{picture}
\nonumber
\ee

\vskip 1.0cm

\noindent In the temporal gauge it corresponds to the expression
\bea
R_{ni,0}
&=& 
\frac{1}{N_{c}} Re ~ Tr [U_i(x) U_i(x+a\hat{i}) 
\cdots  U_i(x+(n-1)a\hat{i}) U_i(x+na\hat{i})
\nonumber \\
&\times& 
U_i^{\dagger} (x+na\hat{i},t+a_t) 
U_i^{\dagger} (x+(n-1)a\hat{i}) 
\cdots U_i^{\dagger} (x+a\hat{i}) 
U_i^{\dagger} (x,t+a_t)].
\nonumber \\
\label{DefRni0}
\eea
Note that for $n=1$, $R_{ni,0}$ coincides with $R_{i0}$. 
The path-ordered integral of such a Wilson loop is given by
\begin{eqnarray}
\oint A  \cdot dx 
&\to& 
\int d x_i dt
[F_{i0}(x_0) +  {1 \over 2} x_{i'}x_{j'} 
({\cal D}_{i'} {\cal D}_{j'} F_{i0}) \vert_{x=x_0}]
\nonumber \\
&=& 
2a a_t F_{i0}(x_0) + 
{1 \over 2} a_t {\cal D}_{i}^2 F_{i0}(x_0) 
[\int_{(n-1)a/2}^{(n+1)a/2}x^2 dx 
+\int_{-(n+1)a/2}^{-(n-1)a/2}x^2 dx]
\nonumber \\
&=& 
2a a_t F_{i0}(x_0) + 
{3n^2+1\over 12} a_t a^3 {\cal D}_{i}^2 F_{i0}(x_0).
\nonumber \\
\end{eqnarray}
Therefore, we obtain the following continuum behavior for the
above Wilson loop
\begin{eqnarray}
R_{ni,0} &\to& \frac{1}{N_{c}} Re ~ Tr [1- {1 \over 2} (\oint A \cdot dx)^2]
\nonumber \\
&\to& 
1 - \frac{g^2 a^{2} a_t^2}{2 N_{c}} 
\left( 4 Tr [F_{i0} F_{i0}]
+ (n^2 + 1/3) a^2 Tr[F_{i0}{\cal D}_{i}^2 F_{i0}] \right).
\nonumber \\
\label{ContRni0}
\end{eqnarray}
One verifies for $n=1$ that Eq.(\ref{ContRni0}) coincides with Eq.(\ref{ContRi0}), as should be.

\bigskip

We make the following ansatz for the Euclidean lattice Lagrangian
\be
L_{t} = - \frac{2 N_{c} a}{g^{2} a_{t}^{2}} A' 
\left[ B' \sum_{x,i} P_{io}(x) + \sum_{n=1}^{\infty} C'^{n} \sum_{x,i} R_{ni,0}(x) \right].
\label{ImprovLagrangian}
\nonumber \\
\ee
In order that the usual continuum limit of the Lagrangian is obtained and the $O(a^{2})$ error is canceled, we imply from the continuum behavior of $P_{i0}$, Eq.(\ref{ContPi0}), and of $R_{ni,0}$, Eq.(\ref{ContRni0}), that the following conditions hold,
\bea 
&& A' [ B' + 4 \sum_{n=1}^{\infty} C'^{n} ] = 1,
\\
&& A' [ \frac{B'}{12} + \sum_{n=1}^{\infty} (n^{2} + \frac{1}{3}) C'^{n} ] = 0.
\eea
We have deliberately introduced the coefficient $B'$. Choosing 
\be
B' = 1 - 2 \sum_{n=1}^{\infty} C'^{n},
\ee
results in a simple expression of the Lagrangian expressed in terms 
of generalized coordinates and velocities. Using 
\begin{eqnarray}
\sum_{n=1}^{\infty} n^2 C'^n
= C' {\partial \over \partial C'} (C' {\partial \over \partial C'}) 
\sum_{n=1}^{\infty}  C'^n
={C' (1+C') \over (1-C')^3},
\end{eqnarray}
we obtain
\bea
&& A' = \frac{1 - C'}{1 + C'},
\nonumber \\
&& B' = \frac{1 - 3 C'}{1 - C'},
\eea
and $C'$ is a root of
\be
C'^3+11C'^2+11C'+1 = 0.
\ee
This equation has three real roots, given by
\bea
C' &=& -1,
\nonumber \\
C'&=&-5 \pm 2 {\sqrt 6}.
\label{SolC'}
\eea
The root closest to zero is 
$C'_{0} = -5 + 2 \sqrt{6} = -0.101021\cdots$. 
In order to obtain the kinetic energy, we express $R_{ni,0}$ in terms of generalized coordinates and velocities,
\begin{eqnarray}
R_{ni,0} 
\nonumber \\
&\to& 
\frac{1}{N_{c}} Re ~ Tr \left[
[U^{\dagger}_i(x)+a_t {\dot U}^{\dagger}_i(x)
+{1\over 2} a_t^2 {\ddot U}^{\dagger}_i(x)] U_i(x) 
U_i(x+a\hat{i}) \cdots \right.
\nonumber \\
&\times&
U_i(x+(n-1)a\hat{i}) 
U_i(x+na\hat{i}) 
[U^{\dagger}_i(x+na\hat{i}) +a_t {\dot U}^{\dagger}_i(x+na\hat{i})
\nonumber \\
&+& \left.
{1\over 2} a_t^2 {\ddot U}^{\dagger}_i(x+na\hat{i})]
U_i^{\dagger} (x+(n-1)a\hat{i}) \cdots U_i^{\dagger} (x+a\hat{i}) \right]
\nonumber \\
&\to&  
\frac{1}{N_{c}} Tr [ 1-{1 \over 2}a_t^2{\dot q}_i(x) {\dot q}_i(x)
-{1 \over 2}a_t^2{\dot q}_i(x+na\hat{i}) {\dot q}_i(x+na\hat{i}) 
\nonumber \\
&-& 
a_t^2 {\dot Q}_i(x+(n-1)a\hat{i}) {\dot q}_i(x+na\hat{i}) ],
\nonumber \\
\label{GenVelRni0}
\end{eqnarray}
where we have introduced
\begin{eqnarray}
{\dot Q}_i(x+(n-1)a\hat{i}) 
\nonumber \\
&=&
U^{\dagger}_i(x+(n-1)a\hat{i}) \cdots U^{\dagger}_i(x+a\hat{i})
U^{\dagger}_i(x) {\dot q}_i(x)U_i(x)
\nonumber \\
& \times &
U_i(x+a\hat{i}) \cdots U_i(x+(n-1)a\hat{i}) . 
\nonumber \\
\end{eqnarray}
Note again, for $n=1$, ${\dot Q}_i(x+(n-1)a\hat{i})$
coincides with ${\dot Q}_i(x)$ defined in Eq. (\ref{Q}). 
Thus we can write the 
time-like part of the Minkowski lattice Lagrangian,
\bea
L^{M}_{t}
= 
{a \over g^2} A'
\sum_{x,i} Tr \left[ {\dot q}_i(x){\dot q}_i(x) + 
2 \sum_{n=1}^{\infty}
C'^n {\dot Q}_i(x+(n-1)a\hat{i}){\dot q}_i(x+na\hat{i}) \right].
\label{GenVelLagr}
\eea
To find the kinetic energy of the improved Hamiltonian,
it is convenient to express this in terms of  
the matrix $J$, defined by Eq.(\ref{DefJ}), 
\be
L^{M}_t={a \over 2 g^2}
\sum_{\sigma \rho} {\dot q}_{\sigma} M'_{\sigma \rho}
{\dot q}_{\rho},
\ee
where
\be
M'= A' \left[1 + 2 \sum_{n=1}^{\infty} C'^n J^n \right]
= A' {1+C'J \over 1-C'J}.
\ee
The color-electric field is expressed as
\be
E_{\sigma} = 
\frac{\partial L } { \partial {\dot q}_{\sigma} }
= \frac{a}{g^{2}} \sum_{\rho} M'^{sym}_{\sigma, \rho} \dot{q}_{\rho}.
\ee
where
\be
M'^{sym}={1 \over 2} (M'+M'^t)  
= A' \frac{ 1 - C'^{2} }{ (1-C'J)(1-C'J^{t}) }
= \frac{ (1-C')^2 }{ (1+C'^2) -C' (J+J^t) }.
\ee
If we choose $C'$ such that $|C'| \neq 1$, e.g. $C'_{0}=-0.101021\cdots$, then $M'_{sym}$ is a real, symmetric, positive and non-singular matrix. 
Finally, we obtain the corresponding kinetic energy of the improved 
Hamiltonian, given by
\begin{eqnarray}
H_{0} &=& \frac{1}{2} \frac{g^{2}}{a} 
\sum_{\sigma, \rho} E_{\sigma} (M'^{sym})^{-1}_{\sigma, \rho} E_{\rho}
\nonumber \\
&=&
{g^2 \over 2 a} \sum_{\sigma, \rho}  
\left[ {1+C'^2 \over (1-C')^2} E_{\sigma} \delta_{\sigma, \rho} E_{\rho}
- {C' \over (1-C')^2} E_{\sigma} (J + J^{t})_{\sigma, \rho} E_{\rho} \right].
\nonumber \\
&=&
{g^2 \over 2 a} \sum_{\sigma, \rho}  
\left[ {1+C'^2 \over (1-C')^2} E_{\sigma} \delta_{\sigma, \rho} E_{\rho}
- {2C' \over (1-C')^2} E_{\sigma} J_{\sigma, \rho} E_{\rho} \right].
\nonumber \\
&=&
{g^2 \over a} Tr \sum_{x,i} 
\left[ {1+C'^2 \over (1-C')^2} E_i(x) E_i(x)
- {2C' \over (1-C')^2} ~ U_i(x)^{\dagger} E_i(x) U_i(x) E_i(x+a\hat{i}) \right].
\nonumber \\
\end{eqnarray}
It consists of only two terms, which makes it convenient for practical calculations.

\section{\bf Tadpole Improvement}
In the preceding section, we derived a classically improved
Hamiltonian for gluons with $O(a^2)$ corrections.
A very important step of the improvement program 
is to take into account quantum corrections corresponding to 
tadpole diagrams. Without such improvement,
only part of the $O(a^2)$ errors are canceled.
According to Lepage and Mackenzie, most of the tadpole
contributions can be removed 
by dividing each link operator $U_{\mu}$ 
by the mean $u_\mu$ of the link.
For asymmetric lattices, $a_t << a_s$,
and small enough $a_t$ we have $u_t = 1$ for time-like directions.
In the Hamiltonian formulation,
the mean $u_s$ of a space-like link
is defined by
\be
u_s=\langle \Omega \vert P_{ij} \vert \Omega \rangle^{1/4},
\ee
where $\vert \Omega \rangle$ 
is the vacuum of the improved Hamiltonian.
Thus tadpole improvement of the lattice Lagrangian
$L = L_t + L_s$, where $L_t$ is given by Eq.(\ref{ImprovLagrangian}) and $L_s$ by Eq.(\ref{ImprovEuclLagr}),
corresponds to the replacements
\bea
P_{ij} &\rightarrow& P_{ij}/u_s^4,
\nonumber \\
R_{ij} &\rightarrow& R_{ij}/u_s^6,
\nonumber \\
P_{i0}& \rightarrow& P_{i0}/u_s^2,
\nonumber \\
R_{ni,0} &\rightarrow& R_{ni,0}/u_s^{2n+2}.
\label{ReplLoops}
\eea
This is equivalent to the following replacement of constants
\bea
C_1 &\rightarrow& C_1/u_s^4,
\nonumber \\
C_2 &\rightarrow& C_2/u_s^6,
\nonumber \\
g_t &\rightarrow& g_t u_s,
\nonumber \\
C' &\rightarrow& C'/u_s^{2},
\label{ReplConst}
\eea
where we put $g=g_t$ in Eq.(\ref{ImprovLagrangian}).
For the transition to the Hamiltonian these redefinitions
of the coefficients can be taken over yielding for the
``two-term'' tadpole improved Hamiltonian 
($C' = -0.101021$)
\bea
H &=& H_0 + V,
\nonumber \\
H_0 &=&
{g_t^2 u_s^2 \over a} Tr \sum_{x,i} 
\left[ {1+C'^2/u_s^4  \over (1-C'/u_s^2)^2 } E_i(x) E_i(x) \right.
\nonumber \\
& - & \left. {2C'/u_s^2  \over (1-C'/u_s^2)^2 } ~ U_i(x)^{\dagger} 
E_i(x) U_i(x) E_i(x+a\hat{i}) \right],
\nonumber \\
V &=&
- {2N_c \over g_s^2 a} \sum_{x,i<j} 
\left[ {C_1 \over u_s^4} {P_{ij}+P_{ji} \over 2} 
+ {C_2 \over u_s^6} (R_{ij}+R_{ji}) \right].
\label{TadImprovHam}
\eea
Here, we have introduced different couplings in the
kinetic and potential terms in order to 
allow for a ``speed of light'' correction as discussed
in Ref.\cite{Hamer96} (see below).

\section{Further perturbative improvement}
Tadpoles have been identified as an essential part of the problem when approaching the continuum limit of quantum field theory on the lattice. 
A systematic perturbative calculation on the lattice 
has been performed by L\"uscher and Weisz\cite{Luscher85}. This 
leads to the determination of additional
terms in the Lagrangian needed to compensate
errors. It turns out that such a further
improved Lagrangian (for details see
Ref.\cite{Lepage}) contains the same
plaquettes and planar rectangle loop terms which occurred before, but with suitably redefined coefficients,
plus a new term, being a non-planar "parallelogram" loop, 
given by  
\be
C_{\mu\nu\sigma}  = { 1 \over N_{c}} Re ~ Tr
\setlength{\unitlength}{.015in}
\begin{picture}(60,20)(0,17)
  \put(10,10){\vector(0,1){12.5}}
  \put(10,10){\line(0,1){20}}
  \put(10,30){\vector(2,1){10}}
  \put(10,30){\line(2,1){15}}
  \put(25.2,37.6){\vector(1,0){12.5}}
  \put(25.2,37.6){\line(1,0){20}}
  \put(45.2,37.6){\vector(0,-1){12.5}}
  \put(45.2,37.6){\line(0,-1){20}}
  \put(45.2,17.6){\vector(-2,-1){10}}
  \put(45.2,17.6){\line(-2,-1){15}}
  \put(30,10){\vector(-1,0){12.5}}
  \put(30,10){\line(-1,0){20}}
\end{picture}.
\ee
It corresponds to 
\be
C_{x,\mu\nu\sigma} = \frac{1}{N_{c}} Re ~ Tr (U_{x,\mu}U_{x+ a\hat\mu,\nu} 
U_{x+a\hat\mu+a\hat\nu,\sigma} U^{-1}_{x+a\hat\nu+a\hat\sigma,\mu} 
U^{-1}_{x+ a\hat\sigma,\nu} U^{-1}_{x,\sigma}).
\label{Parallogram}
\ee
The corresponding term occurring in the Lagrangian is proportional to
\be
\sum_{x,\mu<\nu<\sigma} C_{x,\mu\nu\sigma} . 
\label{ParallLagr}
\ee
The structure of the corresponding improved Hamiltonian
can be inferred from the improved Lagrangian as
before: One introduces different lattice spacings
$a_s = a$ and $ a_t$ and constructs the Hamiltonian
by Legendre transformation and canonical quantization.
Here, we refrain from discussing details and only give
the general structure of emerging Hamiltonian. \\
\noindent (1) The {\em plaquette and planar
rectangle loop} terms will give a part of the improved
Hamiltonian which has  the same form as before, only
the weights of the individual terms will be different. \\
\noindent (2) The {\em space-like parallelogram loop} terms (i.e. 
$\mu\nu\sigma$ space-like) will yield
a corresponding additional term in the potential
part of the Hamiltonian. \\
\noindent (3) The {\em time-like parallelogram loop} terms 
(where either $\mu$ or $\nu$ or  $\sigma$ 
is time-like, the other two indices being space-like) 
produce  a large number of different
contributions to the Hamiltonian (with well defined
weights).
The final result for the improved Lagrangian has the
structure
\be
L = L_t(\dot q, U) + L_s(U),
\ee
with
\be
L_t={a \over 2 g_t^2} \left[
\sum_{\sigma \rho} {\dot q}_{\sigma} M_{\sigma \rho}(U)
{\dot q}_{\rho}
+\sum_{\sigma} A_\sigma(U) \dot q_\sigma + h.c. \right] . 
\label{ParallLt}
\ee
A new feature
is the occurrence of a term linear in $\dot q$.
As before,  $M(U)$ is a symmetric
matrix of the form
\be
M = 1 + \tilde M ,
\ee
allowing the definition of $M^{-1}$ by a geometric series
expansion.
Legendre transformation 
and quantization yields
a Hamiltonian of the structure
\be
H = H_0 + V, 
\ee
with
\be
H_0 = {g_t^2 \over 2a}  
\sum_{\sigma \rho} \left[ E_{\sigma} M^{-1}_{\sigma \rho}
{E}_{\rho} -  (A_\sigma M^{-1}_{\sigma \rho}E_\rho + h.c.)
-2 A_{\sigma} M^{-1}_{\sigma \rho}
A_{\rho}\right].
\label{ParallH0}
\ee

\section{\bf Discussion}
For the purpose of a numerical calculation, in particular for a
comparison with lattice Monte Carlo
results, the following points are important: \\
\noindent (1) As discussed in Refs.\cite{Hasenfratz81,Hamer96}, the scales
related to the regularization of the
gauge field theory in the Hamiltonian formulation as opposed to the
Euclidean path integral formulation are different.
This difference can be accounted for by
introducing spacelike ($g_s$) and time-like
couplings ($g_t$) which have a well defined relation to
the "Lagrangian coupling" $g$. In one-loop 
approximation this relation is of the type
\be
{1 \over g_\mu^2} =  {1 \over g^2} + c_\mu,
\ee
where $c_\mu$ depends on the space-time dimension
and on the type of the gauge group and 
is given in detail in Refs.\cite{Hasenfratz81,Hamer95,Hamer96}. \\
\noindent (2) Because of this difference in the nature of 
the lattice regularization, all perturbative
calculations which determine some non-classical
improvement in the sense of L\"uscher-Weisz
have to be redone. Such a calculation can be done 
on an asymmetric Euclidean lattice 
with $a_t << a_s$ (see Ref.\cite{Hamer96}). \\
\noindent (3) Tadpole improvement which has been considered
by Lepage\cite{Lepage} in the Lagrangian framework corresponds in the Hamiltonian framework to an expression given by Eq.(\ref{TadImprovHam}). \\
\noindent (4) A systematic determination of the 
L\"uscher-Weisz improvement terms on asymmetric
lattices in the Hamiltonian framework
has still to be done. Since these additional
corrections turn out to be small in the
standard Euclidean framework (see Ref.\cite{Lepage}) 
- the most important correction coming from
the inclusion of the tadpole terms -
in should be worthwhile to work with the improved 
Hamiltonian given by Eq.(\ref{TadImprovHam}), e.g., 
for the numerical simulation of glueballs.

\bigskip

To summarize, we have 
investigated in this paper two schemes of improvement
of the Kogut Susskind Hamiltonian:
If one starts from Lepage's Lagrangian, which is preferable for
Monte Carlo simulations in the Lagrangian formulation,
the corresponding Hamiltonian is given by an infinite series of terms which contain terms with arbitrary long range.
In contrast, we have shown that
by starting from a suitable Lagrangian with an infinite number of
terms , one can get an improved Hamiltonian 
consisting of a finite small  number of terms.
This should be preferable for numerical 
computations in the Hamiltonian framework.

\bigskip 

\eject 

\noindent {\bf Acknowledgment} \\
H.K. would like to acknowledge support by NSERC Canada.
X.Q.L. and S.H.G. would like to would like to acknowledge support 
by the Chinese National Natural Science Foundation
and National Education Committee. 
X.Q.L. is grateful to the hospitality offered by the colleagues of 
the Physics Department, Universit\'e Laval, where part of the work
was done. We also thank K. Chao, X. Fang, H. Jirari, T. Huang, J. Liu, Z. Mei, 
N. Scheu, and J. Wu for useful discussions.

\end{document}